\documentclass{emulateapj}
\usepackage{tablefootnote}
\usepackage{threeparttable}
\def\arcmin{{$^{\prime}$}}
\def\arcsec{{$^{\prime\prime}$}}

\def\ptsec{$''\mskip-7.6mu.\,$}
\def\Msun{\,{\rm M$_{\odot}$}}
\def\Lsun{\,{\rm L$_{\odot}$}}

\def\degr{$^{\circ}$}

\shorttitle{FORCAST imaging of two clusters}
\shortauthors{Sandell et al.}

\begin{document}

\title{FORCAST imaging of two small nearby clusters: the Coronet and B\,59}

\author{G\"oran Sandell}
\affil{ Institute for Astronomy, University of Hawai`i at Manoa,  Hilo,  640 N. Aohoku Place, Hilo, HI 96720, USA}
\email{gsandell@hawaii.edu}

\author{Bo Reipurth}
\affil{Institute for Astronomy, University of Hawai`i at Manoa,  Hilo,  640 N. Aohoku Place, Hilo, HI 96720, USA}

%WDV
\author{William D.\ Vacca}
\affil{SOFIA-USRA, NASA Ames Research Center, MS 232-12, Building N232, Rm.
147, PO Box 1, Moffett Field, CA 94035-0001,
USA }

\and     

\author{Naman S. Bajaj}
\affil{Department of Mechanical Engineering, College of Engineering Pune, India}
\begin{abstract}
We present mid infrared imaging of two young clusters, the Coronet in the CrA cloud core and B\,59 in the Pipe Nebula,  using the FORCAST camera on the Stratospheric Observatory for Infrared Astronomy.
We also analyze {\it  Herschel} Space Observatory\footnote{Herschel is an ESA space observatory with science instruments provided by 
European-led Principal Investigator consortia and with important participation from NASA.} PACS and SPIRE images of the associated clouds.
The two clusters are at similar, and very close, distances.  Star formation is 
ongoing in the Coronet, which hosts at least one Class 0 source and several pre-stellar cores, which may collapse and form stars.
The B\,59 cluster is older, although it still has a few Class I sources, and is less compact. The CrA cloud has  a diameter of 
$\sim$ 0.16 pc, and we determine a dust temperature of 15.7 K and a star formation efficiency of  $\sim$ 27 \%, while the B\,59 core is approximately twice as large, has a dust 
temperature of $\sim$ 11.4 K and a star formation efficiency of  $\sim$ 14 \%.  We infer that the gas densities are much higher in the Coronet, which has also formed
intermediate mass stars, while B\,59 has only formed low-mass stars.

\end{abstract}


%\date{latest  draft \today }  
%circumstellar matter, infrared astronomy, 
\section{Introduction}

Most stars form in clusters \citep[e.g.,][]{Zinnecker93}. The clusters range from small
groups of  only a few low mass stars  up to massive OB associations with
thousands of stars. Approximately 75\% of all embedded young stars within 1 kpc
from the Sun are in clusters with 10 or more members \citep[e.g.][]{Lada03}.
\citet{Gutermuth09}  found that a typical cluster core contains 26 members
embedded within a radius of 0.39 pc and has an extinction of 0.8 mag in K. Although
the nearest clusters allow us to study their properties in detail, but they are
often so bright that they are saturated in Spitzer IRAC and MIPS images.
Fortunately, these clusters can be imaged with the FORCAST instrument aboard
SOFIA, which offers better spatial resolution than MIPS and saturates at higher
flux levels. FORCAST is therefore the ideal instrument for studying nearby
clusters.

In this paper we study two nearby clusters, the Coronet in the Corona Australis  (CrA) cloud, 
and the B\,59 cluster in the Pipe Nebula.

The Corona Australis dark cloud complex is a nearby star forming region  forming
low- to intermediate mass stars. The western part of the cloud, where the Herbig
AeBe stars R CrA and T CrA are located, is the most active part of the cloud.
\citet{Taylor84} discovered a small cluster of $\sim 12$ embedded infrared
sources, which they named the Coronet. Of these, only R CrA and T CrA are
visible in the optical.  The Coronet has been studied extensively from radio
wavelengths to X-rays, see e.g. the review by \citet{Neuhauser08}.  The
cloud core in which these sources are embedded is very dense and opaque with a
visual extinction of 50 mag or more \citep{Sicilia11,Bresnahan18} and a mass of
$\sim$ 40 \Msun\ \citep{Harju93} In this dense cloud core are 12 young stellar
objects (YSOs) as well as a prestellar core \citep{Wilking86, Nutter05,
Groppi07,Peterson11}. Almost all the embedded sources in the Coronet are Class I \citep{Groppi07,Chen10,Peterson11,Lindberg14}. 
All of them are X-ray sources \citep{Forbrich06,Forbrich07,Choi08} and all except one,
IRS\,9, are also cm-radio sources \citep{Brown87,Forbrich07,Choi08,Miettinen08}. Many of
them drive outflows \citep{Wang04,Peterson11}. 
The Coronet is at a distance of 154 $\pm$ 4 pc \citep{Dzib18,Mesa19}. The age
of the Coronet cluster is 0.5 -- 1 Myr \citep{Nisini05,Sicilia08}.

The young star cluster B\,59 at the northwestern edge of the Pipe nebula is
also well studied. The cluster is at a distance  of 163 $\pm$ 5 pc
\citep{Dzib18}, i.e., approximately the same distance as the Coronet. The whole
nebula was mapped in CO(1--0) and isotopologues by \citet{Onishi99}, who
identified 14 cloud cores in C$^{18}$O. They found that the B\,59 cloud core was
the only one showing star formation activity. They also found that it was much denser
than the other cores in the Pipe nebula with a mass of $\sim$ 35 \Msun.
\citet{Lombardi06} created a high resolution  extinction map of the whole Pipe
nebula using the 2MASS point source catalogue, while \citet{Roman09} created an
even higher resolution map, which resolved the B\,59 core with a visual extinction $>$ 70 mag.  Analysis of {\it Spitzer} mid-IR
(IRAC and MIPS) data  \citep{Brooke07,Forbrich09} showed that the B\,59 core has
formed a cluster with about 15 YSOs. A follow-up study with moderate resolution
near-infrared spectroscopy confirmed that almost all the IRAC sources are young
pre-main-sequence stars. Contrary to the Coronet, B\,59 is only forming
low-mass stars. About half of the sources in B\,59 are Class II objects, the
others are mostly Class I or flat spectrum sources \citep{Brooke07,Forbrich09}. The
two most deeply embedded objects could be low-mass Class 0 sources 
\citep{Riaz09,Alves19}. Although several of  the B\,59 YSOs are X-ray sources
\citep{Forbrich09,Forbrich10} or cm-radio sources \citep{Dzib13}, it is a much
smaller fraction than in the Coronet. Only three or four  of the YSOs
drive outflows \citep{Duarte12}. The B\,59 cluster appears to be more evolved
than the Coronet cluster.  \citet{Covey10} estimated  an age of $\sim$ 2.6 Myr
for the Class II sources in the cluster.

In this paper we compare the two regions using  unique SOFIA/FORCAST observations
as well as unpublished {\it Herschel} Space Observatory PACS and SPIRE imaging and
photometry.

\section{Observations}
\subsection{FORCAST Observations}
\label{sect-FORCAST}

The Coronet and the B\,59 cluster (Program ID: 07\_0045; PI: B. Reipurth) were
observed with  the Stratospheric Observatory for Infrared Astronomy (SOFIA)
\citep{Young12} on a flight originating from Christchurch, New Zealand on July 8 2019 in very dry
atmospheric conditions using FORCAST on SOFIA.
FORCAST is a dual-channel mid-infrared camera covering the
wavelength region  5 to 40 $\mu$m with a number of filters
\citep{Herter12}. The Short Wave Camera comprises a Si:As array, which is used
for observations below 25 $\mu$m wavelengths, whereas the Long Wave Camera uses
an Si:Sb array for observations longwards of  25 $\mu$m. Both cameras can be used
simultaneously with a dichroic filter. The 256 $\times$ 256 pixel array has a
field of view of 3.4\arcmin{} $\times$ 3.2\arcmin, with a plate scale of
0\ptsec768 per pixel.

The Coronet was observed at an altitude of 39,000 - 39,500 ft on a 107 minute
leg.  The observations were done in the Chop-Offset-Nod (C2NC2) observing mode with
a chop throw of 350\arcsec\ at a Position Angle (P.A.)\footnote{Position Angle is measured 
counterclockwise from north.} = 95\degr\ with a 10\arcmin\ nod at P.A. 155\degr\  
to insure good image quality. We used a 3-point dither with 10 cycles of 60
seconds in the 19.7, 25.3, 31.5 and 37.1 $\mu$m filters.  A few integrations
were excluded in the pipeline processing due to problems with tracking and
vignetting of the telescope. The final integration times were 589, 479, 629, and
375 seconds for the 19.7, 25.3, 31.5, and 37.1 $\mu$m filters, respectively.

B\,59  was observed  at an altitude of 41,900 - 42,000 feet on a  71 minute leg.
These observations were made using symmetric Nod-Match-Chop (NMC) chop  with a
 180\arcsec\ chop at a P.A. of 36\degr. 
Here we also used
a 3-point dither with 10 cycles of  60 seconds for the 11.1 and 31.5 $\mu$m
filter combination, and  15 cycles for the 19.7/37.1 and 25.3/37.1 $\mu$m filter
combinations. Some integrations were lost due to poor tracking and vignetting.
The final integration times were 475, 721, 694, 475, and 1220 seconds for the
11.1, 19.7, 25.3, 31.5, and 37.1 $\mu$m filters respectively.

We retrieved the  pipeline-reduced level 3 data from the SOFIA
archive. These are fully calibrated images, which require no further processing,
(see \citep{Herter13}). The image quality was very good, and diffraction limited at
25.3 $\mu$m and longer wavelengths. For the  19.7 $\mu$m filter there is some broadening due
to the shear layer. For the Coronet we measure Half Power Beam Widths (HPBW)
of 2\ptsec8, 2\ptsec7, 3\ptsec1, and 3\ptsec4 for the 19.7, 25.3, 31.5, and 37.1
$\mu$m filters, respectively. Since our images contain multiple sources with
accurately known coordinates, we were able to assess the accuracy of the astrometry in the FORCAST images. 
We found that the astrometry calibration in the Level 3 images is remarkably good, better than 1\arcsec, 
which is only a fraction of the instrumental FWHM. 
For the Coronet we did make a small, 0\ptsec7, correction using T CrA
as a reference. Photometry of all the images was performed using
APT\footnote{http://www.aperturephotometry.org} and GAIA\footnote{GAIA is a
derivative of the Skycat catalogue and image display tool, developed as part of
the VLT project at ESO. Skycat and GAIA are free software under the terms of the
GNU copyright.}, the latter is part of the STARLINK software suite. Both give
essentially identical results. Whenever possible we used a sky annulus, while for
crowded regions we chose the sky regions to be as representative as possible.
For crowded or faint sources we used small apertures and derived
aperture corrections from a 'curve of growth' method using sources within the image
that were judged to be unresolved. The
average of individual correction factors was then used as the aperture
correction factor for different apertures. Assuming the error in background flux
calculations is the same over the source, we used sky-sigma for obtaining error
in source flux. The photometry for the Coronet region is presented in Table~\ref{tbl-CrAFORCAST}
and for B\,59 in Table~\ref{tbl-B59FORCAST}. The errors are only statistical and do not include calibration
uncertainty, which is about 5\%, or errors in the aperture correction factors, which in most cases 
are less than 5\%.

\begin{deluxetable*}{lllrrrr}[t]
%\tabletypesize{\scriptsize}
\tablecolumns{7}
%\tablenum{1}
\tablewidth{0pt} 
\tablecaption{Positions and flux densities of Coronet sources observed with FORCAST. \label{tbl-CrAFORCAST}}
\tablehead{
\colhead{Source} & \colhead{$\alpha$(2000.0)} & \colhead{$\delta$(2000.0)}  & \colhead{S(19.7 $\mu$m)}& \colhead{S(25.3 $\mu$m)} & \colhead{S(31.5 $\mu$m)} & \colhead{S(37.1 $\mu$m)}  \\ 
   & \colhead{[$^h$  $^m$ $^s$]}& \colhead{[$^\circ$ \arcmin\ \arcsec ]}& \colhead{[Jy]} & \colhead{[Jy]}  & \colhead{[Jy]} & \colhead{[Jy]} 
}
\startdata
T CrA &  19 01 58.79 & $-$36 57 50.3 & 23.4 $\pm$ 0.11 & 30.7 $\pm$ 0.23 &  29.0 $\pm$ 0.15 & 29.3 $\pm$ 0.28  \\
SMM\,2 & 19 01 58.54  & $-$36 57 08.5 & 0.30 $\pm$ 0.05 & 0.79 $\pm$ 0.06 & 1.54 $\pm$ 0.09  & 2.16 $\pm$ 0.11  \\
IRS\,7\,B & 19 01 56.41 & $-$36 57 28.3 &  4.60 $\pm$ 0.13 & 16.8 $\pm$ 0.29 &  29.6 $\pm$ 0.14  & 41.4 $\pm$ 0.80  \\
CrA-24   & 19 01 55.47 & $-$36 56  51.4 & $<$ 0.2\phantom{aa1111}   &  0.30 $\pm$ 0.06 & 0.33 $\pm$ 0.05 &  $<$ 0.5\phantom{aa1111}  \\
IRS\,7\,A & 19 01 55.31 & $-$36 57 22.1 & 19.0 $\pm$ 0.13 & 61.5 $\pm$ 0.28 & 103.1 $\pm$ 0.16 & 146.3 $\pm$ 0.71 \\
R CrA   & 19 01 53.63 & $-$36 57 07.9 & 181.5 $\pm$ 0.15 & 203.0 $\pm$ 0.23 &  195.0 $\pm$  0.25 & 175.9 $\pm$ 0.61 \\
IRS\,1\tablenotemark{a}  & 19 01 50.70 & $-$36 58 10.6 & 46.8 $\pm$ 0.13 & 81.8 $\pm$ 0.16 &94.7 $\pm$ 0.12 & 104.1 $\pm$ 0.29    \\
IRS\,6 & 19 01 50.47 & $-$36 56 37.7 & 0.35 $\pm$ 0.04 &$<$ 0.7\phantom{aa1111}  & 0.69 $\pm$ 0.07& $<$ 0.8\phantom{aa1111}   \\
IRS\,5\,N& 19 01 48.44 & $-$36 57 14.5 & 0.19 $\pm$ 0.05 & 0.68 $\pm$  0.07 & 1.24 $\pm$ 0.10 & 2.22 $\pm$ 0.14 \\
IRS\,5 &  19 01 48.00 & $-$36 57 22.0 & 5.15 $\pm$ 0.08 & 8.65 $\pm$ 0.12 & 9.45 $\pm$  0.11 & 10.0 $\pm$ 0.21  
 \enddata
 \tablenotetext{a}{ The FORCAST position for IRS\,1 measured relative to other stars in the cluster differs by $\sim$ 1\arcsec\ from the Simbad position. }
\end{deluxetable*}

\begin{deluxetable*}{lrrrrrrr}[b]
%\tabletypesize{\scriptsize}
\tablecolumns{8}
%\tablenum{2}
\tablewidth{0pt} 
\tablecaption{Positions and flux densities of B\,59 sources observed with FORCAST
\label{tbl-B59FORCAST}}
\tablehead{
\colhead{Source}& \colhead{$\alpha$(2000.0)} & \colhead{$\delta$(2000.0)}  & \colhead{S(11.1 $\mu$m)}& \colhead{S(19.7$\mu$m)} & \colhead{S(25.3 $\mu$m)} & \colhead{S(31.5 $\mu$m)} & \colhead{S(37.1 $\mu$m)}   \\ 
   & \colhead{[$^h$  $^m$ $^s$]}& \colhead{[$^\circ$ \arcmin\ \arcsec ]}& \colhead{[Jy]} & \colhead{[Jy]}  & \colhead{[Jy]} & \colhead{[Jy]}  & \colhead{[Jy]} 
}
\startdata
$[$BHB2007$]$\,6 & 17 11 16.39 & $-$27 25 15.2 & 0.25 $\pm$ 0.04 & 0.50 $\pm$ 0.07 & 0.35 $\pm$ 0.10 & 0.17 $\pm$ 0.06 & 0.28 $\pm$ 0.10  \\
$[$BHB2007$]$\,7 & 17 11 17.35 & $-$27 25 08.5 & 3.4\phantom{0} $\pm$ 0.07 & 9.74 $\pm$ 0.08 & 11.07 $\pm$ 0.13 & 10.82 $\pm$ 0.11 & 11.64 $\pm$ 0.17 \\
$[$BHB2007$]$\,8  & 17 11 18.32 & $-$27 25 49.2 & 0.16 $\pm$ 0.04 & 0.37 $\pm$ 0.06 & 0.33 $\pm$ 0.10 & 0.32 $\pm$ 0.08 & 0.54 $\pm$ 0.09 \\
$[$BHB2007$]$\,10 & 17 11 22.26 & $-$27 26 01.9 & $<$ 0.27\phantom{aa1111} & 1.49 $\pm$ 0.09 & 4.49 $\pm$ 0.13 & 6.44 $\pm$ 0.14 & 6.56 $\pm$ 0.16 \\
$[$BHB2007$]$\,11 & 17 11 23.19 & $-$27 24 32.8 & $<$ 0.27\phantom{aa1111} & 0.50 $\pm$ 0.06 & 3.34 $\pm$ 0.13  & 7.79 $\pm$ 0.14 & 10.46 $\pm$ 0.17\\
$[$BHB2007$]$\,13 & 17 11 27.11  & $-$27 23 48.7 & $<$ 0.27\phantom{aa1111} & 0.68 $\pm$ 0.07 & 0.95 $\pm$ 0.09  & 1.11 $\pm$ 0.13 & 1.05 $\pm$ 0.10 \\
$[$BHB2007$]$\,14 & 17 11 27.49 & $-$27 25 28.6 & 0.16 $\pm$ 0.05 & 0.44 $\pm$ 0.05 &  0.40 $\pm$ 0.05 & 0.55 $\pm$ 0.07 & 0.40 $\pm$ 0.10

 \enddata
\end{deluxetable*}

\subsection{Herschel Archive data}
\label{sect-Herschel}

All the PACS data for the Coronet that we have retrieved from the Herschel data archive are fully reduced
level 2.5 or level 3 JScanam images processed  with SPG v14.2.0.  For SPIRE  we used
point source calibrated maps  for photometry.
For  the  PACS 70 $\mu$m data we used the  map observed in parallel mode with
fast scanning  (1342206677, 1342206678), 
program ID (KPGT\_pandre\_1) observed on 2010-10-17 (OD = 521). Photometry from these data sets have been
published by \citet{Sicilia13}. However, since the angular resolution is degraded in 
fast scanning, we additionally made use of  a 70 $\mu$m image observed in 
small map mode  from the program OT1\_gmeeus\_1,
which targeted TY CrA north of the Coronet (1342267427, 1342267427, OD 1400) observed on 2013-03-13. It 
covers only the northernmost portion of the Coronet and therefore R CrA, IRS\,6, and IRS\,5/5 N are seen in this image.

For 100 $\mu$m and 160 $\mu$m there are high quality images from program
OT2\_jforbrich\_3, where the Coronet was observed  in small map mode five times
between 2012-03-13  and 2012-03-28 (OD 1034, 1035, 1037,1041, 1049) with the
intent of looking for far-infrared variability.  There are no indications of
variability and, since the maps are not signal-to-noise limited, we 
performed photometry only on the images from the first epoch. Our photometry (Table
2) agrees well with the measurements generated by the automated photometry on the pipeline-reduced data for the few
sources that were identified by the pipeline software. For blended sources our photometry is
more accurate than the photometry published by \citet{Sicilia13}.

B\,59 was observed in PACS/SPIRE parallel mode under program ID  KPGT\_pandre\_1
on 2010-09-04. 
The PACS  blue channel  was set to 70 $\mu$m for these parallel
observations.  The  AOR IDs are 1342042048 and 1342043049. B\,59 was
additionally observed with PACS at 100 $\mu$m and 160 $\mu$m with medium scan
speed on 2010-09-20. The AOR IDs for these observations are 1342228966 and
1342228967. For the PACS photometry we only use the parallel observations for 70
$\mu$m, since the medium scan speed observations have better image quality at
160 $\mu$m.

\begin{deluxetable*}{lcccccrcc}[h]
\tabletypesize{\scriptsize}
\tablecolumns{9}
%\tablewidth{0pt} 
\tablecaption{Basic properties of the Coronet sources  \label{tbl-CrAbasic}}
\tablehead{
\colhead{Source} & \colhead{Sp. type} &  \colhead{L$_{bol}$} &\colhead{A$_V$} & \colhead{$\alpha$}    & \colhead{IR Class} &\colhead{$\alpha^*$}  &  \colhead{L$_{bol}$} &  \colhead{T$_{bol}$} \\ 
\colhead{}  &  \colhead{} & \colhead{[\Lsun]} &\colhead{[mag]}  &   \colhead{}  & \colhead{}  & \colhead{}    & \colhead{[\Lsun]} & \colhead{[K]}  }

\startdata
T CrA & F0 & 3.6 & 2.45 &-0.35 & n.a. &0.15  & 7.27 (0.36) & 921 (7)  \\    % type I , i.e. flared disk                                    
SMM\,2 &  & 9.7 & 20 & 0.88  & I &0.89 & 0.68 (0.05) & 164 (2)  \\  %Lbol/Lsubmm = 25.9    ;photometry errors only 1.6 K
IRS\,7\,B & &2.0 & 20 & 2.78 & I &  2.04   & 6.2 (0.6)  & 122 (4)  \\  %Lbol/Lsub  =206.3  175 - 242
IRS\,7\,A &  &0.62 & 20 &  2.64 & I &1.67  & 14.9 (0.5)  &  125 (0.3) \\  %Lbol/Lsubmm = 1164 
SMM\,1\,C &  & \nodata & 50 & \nodata & 0 & \nodata & 5.1 (1.6) & 26.8 (0.6) \\
R CrA   & B8 - A5 & 72 &1.33 & -2.55 & n.a.  & -0.45&90.8 (8.1)  & 986 (35) \\   % type II, flat disk
IRS\,1 & K5 - M0 & 21 & 30 & 0.92 & I & 0.19   &18.2 (0.7)  & 418 (12)  \\  %Lbol/Lsubmm =  434   (636 -- 322)
IRS\,6& M1 & 0.15 & 29  &  -0.39 & II & -0.84  & 0.50 (0.01) &  725 (20) \\
IRS\,5\,N&  & 5.2 & 20 & 1.41 & I & 0.33 & 1.01 (0.18)& 133 (11) \\  %  Lbol/Lsubmm = 16.8   9.8 to 31.2
IRS\,5 & K4 & 6.2 & 30 & 0.78 & I &  0.19  & 2.08 (0.02)  & 445 (10)     % 
%CrA-24  0.60 I
\enddata
\tablecomments{Spectral types, when given, come from:  \citet{Finkenzeller84} (T
CrA), \citet{Bibo92} and \citet{Chen97} (R CrA),  \citet{Nisini05} (IRS1) and
\citet{Meyer09} (IRS\,5 and IRS\,6).  The bolometric luminosites, L$_{bol}$,  are
from Dunham et al. (2015) and corrected for distance. For most sources they differ
significantly from our results. We have much better agreement with
\citet{Lindberg14a}, but they only observed a few sources. The visual 
extinction A$_V$ comes from \citet{Acke04} for R CrA and T CrA, and  from
\citet{Nisini05} for IRS\,1 and IRS\,6,  all the rest of the values are our best
estimates for total extinction. The line of sight extinction, used for deriving T$_{bol}$, is assumed to be
half of the total extinction. The spectral index, $\alpha$, and the IR Class are
from \citet{Peterson11}. The last three columns give results from this paper.
$\alpha^*$ is  the extinction corrected spectral index from K' - 19.7 $\mu$m. }
% \tablenotetext{b}{IR Class from \citet{Peterson11}}
% \tablenotetext{c} {Luminosities from Dunham et al. (2015) based on the c2d data presented by \citet{Peterson11}}
% (A$_V$ = 7.9 mag for all sources except IRS\,6 (28 mag). 
%  4 Msun for the T CrA binary
\end{deluxetable*}
% Evans et al.  Tbol <  70 K Class 0;  70 <  Tbol <  650 Class I, 650  <  Tbol <  2800 K Class II

\begin{deluxetable*}{lcccccrcc}[ht]
\tabletypesize{\scriptsize}
\tablecolumns{9}
%\tablewidth{0pt} 
\tablecaption{Basic properties of the B\,59 sources.  \label{tbl-B59basic}.} 
\tablehead{
\colhead{Source} & \colhead{Sp. type} &  \colhead{L$_{bol}$} &\colhead{A$_V$} & \colhead{$\alpha$}    & \colhead{IR Class} &\colhead{$\alpha^*$}  &  \colhead{L$_{bol}$} &  \colhead{T$_{bol}$}   \\ 
\colhead{}  &  \colhead{} & \colhead{[\Lsun]} &\colhead{[mag]}  &   \colhead{}  & \colhead{}  & \colhead{}    & \colhead{[\Lsun]} & \colhead{[K]}  
}
\startdata

$[$BHB2007$]$\,1 & K7 & $>$2.5 & 1.9\tablenotemark{a} & -0.35 & F/F & -0.04\tablenotemark{b}  &2.88 (\phantom{0}0.09) & 900 (35)   \\
$[$BHB2007$]$\,6& M2 & 1.5 &5.5 & -1.14  & II/n.a. & -1.10    &0.47 (0.02) & 1367 (42)  \\
$[$BHB2007$]$\,7 & K5 & 4.2 & 16.0 & 0.02 & F/I & 0.29    &2.97 (\phantom{0}0.08)  & 493 (3.6)  \\
$[$BHB2007$]$\,8 & M & 2.2 & 26.1& -1.04& II/I  & -0.09    &0.17 (0.02) & 570 (20) \\
$[$BHB2007$]$\,9 & K5 &$>$1.4& 7.7& -0.22 & F/F & -0.33\tablenotemark{b}  & 1.08 (0.04)   &  770 (33)   \\
$[$BHB2007$]$\,10 & M & 1.0 & 35\tablenotemark{c}& \nodata &n.a./I & 2.30\tablenotemark{d}     &1.06 (0.04)  & 109 (\phantom{0}1) \\
$[$BHB2007$]$\,11 & M & 3.5 & 20\tablenotemark{c} & \nodata& n.a./I &  1.52    &4.43 (0.04)  & 61.5 (0.5) \\
$[$BHB2007$]$\,13 & M2 & 0.8 & 7\tablenotemark{c}  & -1.04  & II/F& -0.56     &  0.51 (0.03)& 1150  (25) \\
$[$BHB2007$]$\,14 & K5 & 1.3 &14.2& -1.36  & II/II &  -0.87     &0.35 (0.03)& 1420 (40) 
\enddata
\tablecomments{Spectral types and extinctions are from \citet{Covey10} unless noted otherwise.  The bolometric luminosity is from \citet{Brooke07}, as well as the extinction corrected  spectral index, $\alpha$ determined between K$_s$ and IRAC 8 $\mu$m, the first value of the IR class is also from \citet{Brooke07} while the second value is from Forbrich et al. (2009) revised for the spectral index between K$_s$ and 24 $\mu$m. The last three columns give results from this paper. $\alpha^*$ is  the extinction corrected spectral index from K' - 11.1 $\mu$m. }
\tablenotetext{a}{Extinction from \citet{Zurlo21}}
\tablenotetext{b}{Extinction corrected spectral index from K' - 24 $\mu$m}
\tablenotetext{c}{Our best estimate}
\tablenotetext{d}{Extinction corrected spectral index from 3.6 $\mu$m - 19.7 $\mu$m}

\end{deluxetable*}

\section{Results and Analysis}
\label{sect-Results}

The basic properties of the embedded sources  are summarized in
Table~\ref{tbl-CrAbasic} and Table~\ref{tbl-B59basic}. These tables include
parameter values we derive below in order to facilitate comparisons with
previous results.  Spectral Energy Distribution (SEDs) of all sources detected
in the mid-IR are shown in the Appendix, Figure~\ref{fig-Coronet_SEDs} and
\ref{fig-B59_SEDs} The IR classification of the embedded sources in
Tables~\ref{tbl-CrAbasic} and \ref{tbl-B59basic}  is based on spectral indices,
usually from 2MASS K$_s$ to IRAC 8 $\mu$m and corrected for extinction. To
refine the spectral indices of our sources we also compute the extinction
corrected spectral indices from 2MASS K$_s$ to the shortest FORCAST band, i.e.
11.1 $\mu$m for B\,59 and 19.7 $\mu$m for the Coronet. Here we adopt the
extinction law given by \citet{McClure09}. We also draw these spectral slopes as
a red dotted line in the SED plots (Figure~\ref{fig-Coronet_SEDs} and
\ref{fig-B59_SEDs}).  These indices are also given in Table~\ref{tbl-CrAbasic}
and Table~\ref{tbl-B59basic} in the column labelled as $\alpha^*$. If the source
is not detected at 11.1 or  19.7 $\mu$m  we go to the closest band where it is
detected. Both the 11.1 $\mu$m  and the 19.7 $\mu$m can be affected by the
silicate bands at 9.5 $\mu$m and 18 $\mu$m. This is seen quite well in the ISO
high resolution 5 - 30 $\mu$m spectrum of IRS\,1 in the Coronet \citep{Gibb04}.
Although the spectral index can change by a few tenths of a mag, depending on
the exact wavelength range used, for sources with extinctions of more than 20
mag, it is clear that all the sources are Class 0 or I. Only  IRS\,1, which has an
extinction of 30 mag, moves from Class I to a Flat spectrum source.

\begin{figure*}[h]
\includegraphics[width=\textwidth,angle=0]{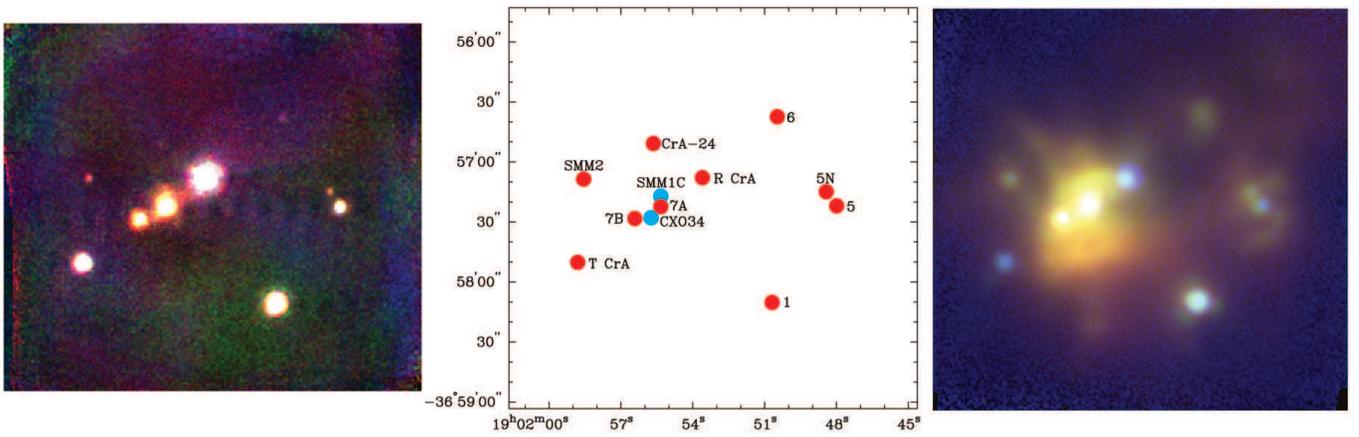}
\figcaption[]{
\label{fig-FORCAST}
The left panel shows a  three-color image made from the FORCAST  19.1 $\mu$m
(blue), 25.3 $\mu$m (green), and 31.5 $\mu$m (red) images. None of the stars
show any extended dust emission. IRS\,7\,A and B are both very red. The
fluctuation of the background are due to imperfections in the flat fielding. In
the middle panel we have marked all the sources detected by FORCAST and PACS
with red filled circles and labeled them. Hence one can easily identify what source
is which when looking at the  color images. For all the IRS sources, we have
dropped the prefix IRS and just label them with the number, i.e., source 1 is
the same as IRS\,1. Additionally we marked the two sub-millimeter sources from
Lindberg et al. (2014) with blue filled circles; SMM\,1\,C and  CXO\,34, the latter is
also detected in the mid-IR but not detected by FORCAST. Since it is close to
the strong source IRS\,7\,B, it is not possible to determine its flux density in
the FORCAST images. At 100 $\mu$m and longer wavelengths SMM\,1\,C dominates
over IRS\,7\,B.  The right panel shows a three-color FORCAST/PACS image. Here
the FORCAST 37.1 $\mu$m image is coded blue, the PACS 100 $\mu$m image green and
the PACS 160 $\mu$m image red.  SMM\,1\,C and IRS\,7\,A blend together creating a source which appears
extended in the north-south direction. Both R CrA and T CrA appear blue in this color image.}    
\end{figure*}

We determine the bolometric luminosity, L$_{bol}$, by directly integrating the
observed SED between the shortest and longest wavelength data points, using a
trapezoidal rule, and multiplying by $4 \pi d^2$. Uncertainties were estimated
for each source by adding (and subtracting) the errors on each photometric data
point to the reported value and re-computing the integral. In addition to the
photometry derived in this paper we also include 2MASS, IRAC and MIPS data,
optical photometry from SIMBAD's Vizier SED plotter and published
millimeter/submilimeter photometry, when available.  Since all our sources are
deeply embedded, the mid infrared photometry generally dominates the bolometric
luminosity. In some cases the published bolometric luminosities differ significantly from our results.
For B\,59 (Table~\ref{tbl-B59basic}) ) the difference is largely due to the more accurate and extensive 
wavelength coverage of the photometry we present in this paper.
For the Coronet (Table~\ref{tbl-CrAbasic}) the \citet{Dunham15} results are inconsistent with the photometric values. 
We note that their extinction 
corrected bolometric luminosities are almost the same as the uncorrected ones, 
yet almost all sources were assigned a visual extinction of 7.9 mag.

The bolometric temperature, $T_{bol}$, is  computed using the
formula 

$$ T_{bol} = 1.25\times 10^{-11} \frac{c \int_{\lambda_{\rm
min}}^{\lambda_{\rm max}} \frac{F^c_{\lambda}}{\lambda}
d\lambda}{\int_{\lambda_{\rm min}}^{\lambda_{\rm max}} F^c_{\lambda} d\lambda}$$

where $F^c_{\lambda}$ is the extinction corrected flux at each wavelength
$\lambda$  \citep{Myers93}.  The bolometric temperature should only be corrected
for line of sight extinction. Since one usually measures the total source
extinction, which includes contribution from the disk and the envelope, we take
the line of sight extinction to be half of the total extinction, which appears
reasonable in view of the
uncertainties involved.

\subsection{The Coronet cluster}

\begin{figure*}
\includegraphics[width=13.5cm, angle=-90]{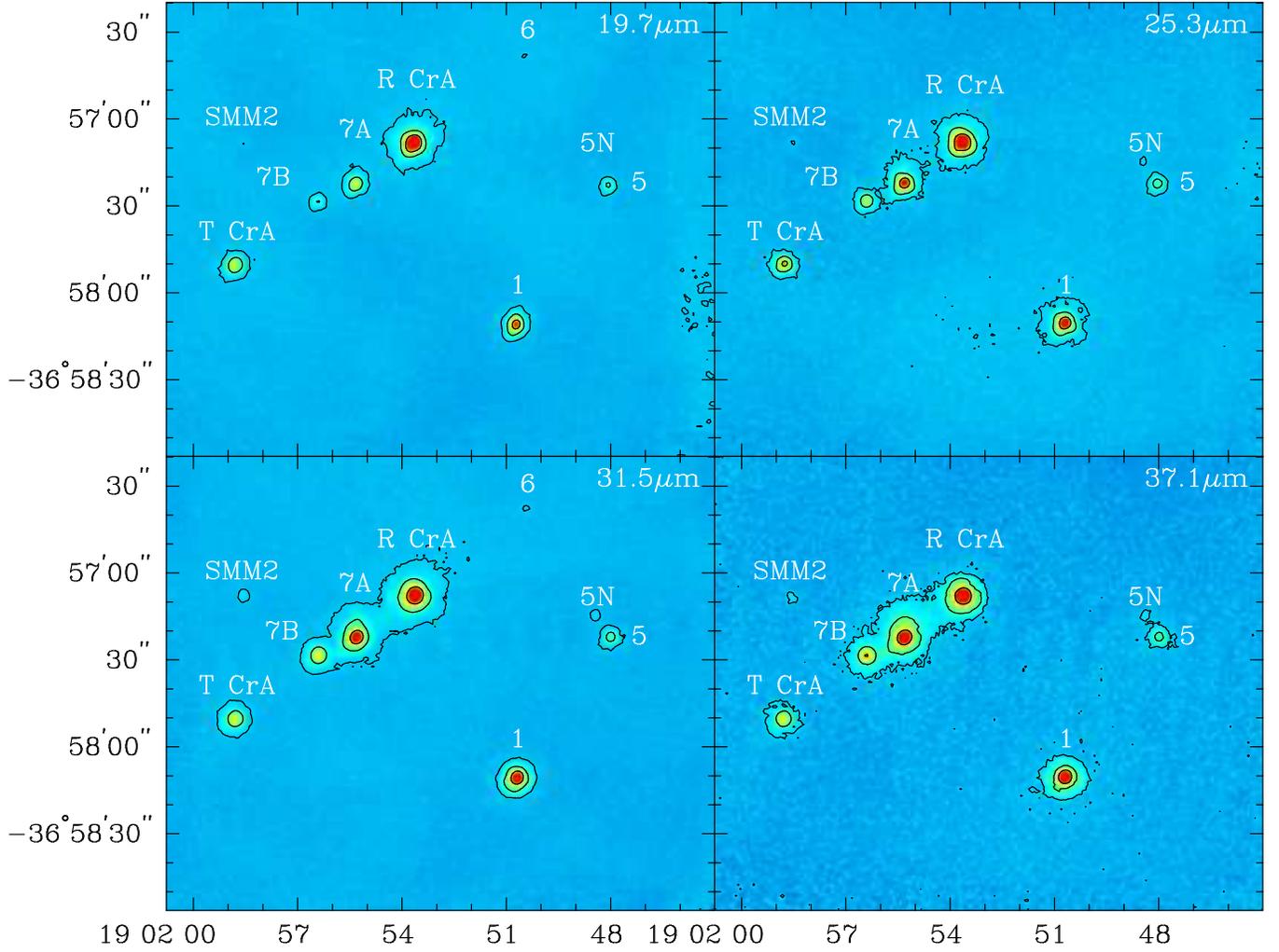}
\figcaption[]{
\label{fig-CrA_4panel}
False color images of the Coronet in the four FORCAST filters, 19.7 $\mu$m, 25.3 $\mu$m, 31.5 $\mu$m and 37.1  $\mu$m. 
The images are plotted with a logarithmic stretch and overlaid with contours to show the low level emission. The lowest
contour is at  about  five sigma rms. All the sources, which are detected, are labelled. At 31.5 $\mu$m and 37.1 $\mu$m the
images show a ridge of hot dust emission  extending to the southeast
from R CrA, through IRS\,7\,A and IRS\,7\,B. Neither CXO\,34 and SMM\,1\,C, are seen in any of the FORCAST images, although both of them are
embedded in the dust ridge, see {\bf central panel in} Figure~\ref{fig-FORCAST}.
 }    
\end{figure*}

The FORCAST images reveal ten sources, all of which were previously known
(Figures~\ref{fig-FORCAST} and \ref{fig-CrA_4panel}, Table~\ref{tbl-CrAFORCAST}). If we exclude the two Herbig AeBe stars all
sources, except IRS\,6, are either Class I or Class 0 (Table~\ref{tbl-CrAFORCAST}), suggesting that they are very young. We did not detect IRS\,9, west of R CrA, nor was it
detected by {\it Spitzer}/IRAC  \citep{Peterson11}.  
At 31.5 and 37.1 $\mu$m one can see a ridge of dust emission
encompassing the two deeply embedded Class I sources IRS\,7\,B and IRS\,7\,A and
extending up to R CrA. CXO\,34, a Class I source between IRS\,7\,A and
IRS\,7\,B, is not apparent in the FORCAST images (Figure~\ref{fig-CrA_4panel}).  All
sources detected by FORCAST are detected by PACS as well except CrA-24, which was barely detected by FORCAST at 25 and 31 $\mu$m.
At SPIRE wavelengths it is too difficult  to extract photometry of the
embedded Coronet cluster members, partly due to the poorer spatial resolution compared
to FORCAST and PACS, but also because the sources are fainter and the emission
from the surrounding cloud is very strong.

\subsubsection{R CrA and T CrA}
R CrA and T CrA are HAeBe stars with strong infrared excesses. They are both highly
variable in the optical. The spectral type of R CrA is uncertain, with classifications 
between late B and early to mid A reported \citep{Bibo92,Chen97}.   R
CrA is most likely a triple system \citep{Sissa19}. It has  a central compact
binary of two intermediate mass stars of roughly equal mass (3 and 2.3 \Msun{})
and an M type tertiary with a separation of 0\ptsec156 from the binary \citep{Mesa19}. T CrA has a spectral type of F0
\citep{Finkenzeller84}. It is also a close binary, detected
using spectro-astrometry, which suggests that the secondary is much fainter than the
primary \citep{Bailey98,Takami03}. Both stars are unresolved with FORCAST. They
have very strong excesses in the FIR. T CrA was detected in  all PACS bands
(Table~\ref{tbl-CrAPACS}). R CrA, which is much brighter at 70 and 100 $\mu$m than T CrA, is no
longer visible at 160 $\mu$m, partly because of the strong emission from the
surrounding cloud (Figure~\ref{fig-FORCAST}).  R CrA was  detected in continuum at 1.3 mm
\citep{Chen10,Peterson11} with the SMA, but it was not detected with SCUBA \citep{Groppi07}. The
emission is very faint, 36 $\pm$ 25 mJy. R CrA is a cm-radio source
\citep{Feigelson98,Forbrich06,Choi08}. Since the radio emission, which is
variable,  has a positive spectral index, $\alpha$ $\geq$ 1 \citep{Choi08} it is
likely that the mm spectrum is due to free-free emission, and not dust emission.
At 1.3 mm the free-free emission is predicted to be $\sim$ 6 mJy, which is
roughly within the 1-$\sigma$ error of the observed flux density. Even though
both R CrA and T CrA  have strong infrared excess from warm or hot dust, they
appear to be devoid of cold gas. In this sense they differ from lower mass stars,
where cold circumstellar disks are ubiquitous and often quite massive, which is
true for all the embedded Class I sources in the Coronet.
% We derive bolometric luminosities
%of 76 and 6.8 \Lsun\ for R CrA and T CrA, respectively.

\begin{deluxetable*}{lllrrr}[h]
\tablecolumns{6}
\tablewidth{0pt} 
\tablecaption{Flux densities of Coronet sources observed with PACS  \label{tbl-CrAPACS}}
\tablehead{
\colhead{Source} & \colhead{$\alpha$(2000.0)} & \colhead{$\delta$(2000.0)}  & \colhead{S(70 $\mu$m)}& \colhead{S(100 $\mu$m)} & \colhead{S(160 $\mu$m)}  \\ 
\colhead{}  & \colhead{[$^h$  $^m$ $^s$]}& \colhead{[$^\circ$ \arcmin\ \arcsec ]}& \colhead{[Jy]} & \colhead{[Jy]}  &  \colhead{[Jy]} 
}
\startdata
T CrA &  19 01 58.83 & $-$36 57 50.6 & 19.3 $\pm$ 0.5 & 14.2 $\pm$ 0.6 & 5.0  $\pm$ 3.0    \\
SMM\,2 & 19 01 58.55  & $-$36 57 08.0 & 8.8 $\pm$  0.8 & 13.5 $\pm$ 0.5 & 13.3 $\pm$ 1.1    \\
IRS\,7\,B & 19 01 56.42 & $-$36 57 29.2 &  99.4  $\pm$ 5.1 &  114.0 $\pm$ 1.5 & 63 $\pm$ 20 \\
IRS\,7\,A & 19 01 55.44  & $-$36 57 21.8& 320.0 $\pm$ 7.6 & 130 $\pm$  40 \tablenotemark{a}& 15 $\pm$ 10\tablenotemark{a}  \\
SMM\,1\,C & 19 01 55.29 & $-$36 57 17.0 & $<$ 10\phantom{aa1111} & 100 $\pm$ 40 \tablenotemark{a} & 90 $\pm$ 30 \tablenotemark{a}\\
R CrA   & 19 01 53.94 & $-$36 57 09.9 & 101.5 $\pm$ 0.7 & 92.5 $\pm$ 2.6 & $<$ 30\phantom{aa1111}   \\
IRS\,1 & 19 01 50.78 & $-$36 58 10.3 & 120.3 $\pm$ 0.2 &  94.6 $\pm$ 0.2 & 40.5 $\pm$ 1.7    \\
IRS\,6\tablenotemark{b} & 19 01 50.48 & $-$36 56 37.8 & 9.0 $\pm$ 2.0  & 13.3 $\pm$ 1.5 &  9.3 $\pm$ 2.0  \\
IRS\,5\,N& 19 01 48.45 & $-$36 57 14.6& 12.7 $\pm$ 0.1 & 19.0 $\pm$ 0.1  & 22.0 $\pm$ 0.5\tablenotemark{c}   \\
IRS\,5 &  19 01 48.15& $-$36 57 23.0 & 20.4 $\pm$ 0.1&  19.7 $\pm$ 0.1 & 22.0 $\pm$ 0.5\tablenotemark{b}  
 \enddata
  \tablenotetext{a}{Blend with SMM\,1\,C, emission extended. }
   \tablenotetext{b}{Emission extended at all PACS wavelengths}
 \tablenotetext{c}{The emission at 160 $\mu$m peaks halfway between 5 \& 5\,N and includes both}
\end{deluxetable*}

\subsubsection{IRS\,1}

IRS\,1 is a mid-IR bright  Class I source which was first discovered as an IR
source illuminating the (mostly reflection) nebula HH\,100 \citep{Strom74}. It is still
very bright in the far infrared (Table~\ref{tbl-CrAPACS}). It drives an outflow and is detected in the sub-millimeter with
SCUBA at both 850 $\mu$m and 450 $\mu$m \citep{Nutter05,Groppi07}. \citet{Nisini05} determined a spectral type of K5 - M0V from medium resolution
IR spectroscopy resulting in a visual extinction  of 30 $\pm$ 3 mag.
 The spectral index and  bolometric temperature that we derive (Table~\ref{tbl-CrAbasic})
suggest that it is a Flat spectrum source. \citet{Chen93} obtained medium-resolution spectroscopy from 2.8 to 3.8
 $\mu$m of  11 embedded sources and eight background stars in the direction of the CrA molecular cloud. They found that 
 the water ice absorption feature at 3.1 $\mu$m was  very prominent for the embedded sources in the cloud. For IRS\,1 their observations
 indicate a line of sight absorption of about 15 mag. This agrees well with the results by \citet{Nisini05}, 30 mag, if our assumption that half of
 the extinction is along the line of sight is adopted.
%  because the outflow from IRS\,1 indicates that 
%  we see the disk surrounding IRS\,1 largely edge on.

\subsubsection{IRS\,7\,A, IRS\,7\,B, and SMM\,1\,C (B\,9)}
\label{sect-SMM1C}

IRS\,7\,A and IRS\,7\,B are two deeply embedded Class I sources close to the
center of the cloud. IRS\,7 was first detected as a nebulous near-IR  object by
\citet{Taylor84}. \citet{Brown87} found two compact radio sources positioned on
each side of the infrared source separated by 14\ptsec2 and suggested that they
are either two separate stars or that the radio emission represents the
shock-excited inner portions of a large accretion disk. \citet{Wilking97} found
that the western radio source coincides with a 10 $\mu$m object and
\citet{Choi04} and \citet{Forbrich06} discovered  that both of the radio sources
found by  \citet{Brown87} have X-ray counterparts, thus securely identifying
them as YSOs. These are now known as IRS\,7\,A (or 7\,W) and IRS\,7\,B (or
7\,E). Both stars were detected with both IRAC and MIPS
\citep{Groppi07,Forbrich07,Peterson11}. At 37 $\mu$m there is a ridge of
emission connecting IRS\,7\,B with IRS\,7\,A continuing up to R CrA
(Figure~\ref{fig-CrA_4panel}). \citet{Choi04} also found an X-ray source
centered on the radio source B\,9 \citep{Brown87}. This source, aka  SMM\,1\,C,
is a Class 0 source, which is so deeply embedded that it is not detected with
IRAC and MIPS \citep{Peterson11}. We do not detect it with FORCAST either
(Table~\ref{tbl-CrAFORCAST}). It appears to be present at 100 and 160 $\mu$m
(Table~\ref{tbl-CrAPACS}), but it cannot be separated from the much stronger emission
from IRS\,7\,A, which is only 4\ptsec4 south of it (Figure~\ref{fig-FORCAST}).
All three sources have been detected in sub-mm continuum with ALMA
\citep{Lindberg14}, although IRS\,7\,A  only marginally at 870 $\mu$m. It was
not detected with SCUBA at 450 $\mu$m \citep{Groppi07}. \citet{Lindberg14}
detected a disk with Keplerian rotation in C$^{17}$O(3-2) toward IRS\,7\,B, from
which they derived a stellar mass of  2.3 \Msun\ assuming an inclination of
60\degr. %relative to the plane of the sky From their continuum observations
they estimated the mass of the disk as 0.034 \Msun\footnote{corrected to d = 154
pc}. Using a least squares two-component graybody fit to the mid and far 
infrared data  from the present paper supplemented with millimeter and
sub-millimeter data from \citet{Groppi07,Peterson11,Lindberg14}, we determine a
dust emissivity index, $\beta$ in the range 1.4 to 1.5 , and a dust temperature
of 39 - 44 K for the cold dust emission,  resulting in a disk mass of 0.045
$\pm$ 0.003 \Msun. The warm gas has a temperature about 80 K and a negligible
mass. Our mass estimate is similar to what \citeauthor{Lindberg14} derived from
their ALMA continuum observations alone, suggesting that the dust emission from
the  disk dominates over  the emission from the envelope. The bolometric
luminosities and temperatures for IRS\,7\,B and IRS\,7\,A are given in
Table~\ref{tbl-CrAFORCAST}, confirming that they are both Class I sources. It is
difficult to estimate the mass and luminosity of SMM\,1\,C, because it is only
detected in X-ray, radio, sub-millimeter, and in the far infrared. In the far
infrared SMM\,1\,C is severely blended by IRS\,7\,A  resulting in highly
uncertain flux densities for both sources (Table ~\ref{tbl-CrAPACS}). Nevertheless,
there is no doubt that SMM\,1\,C is a  Class 0 source. We derive a bolometric
temperature of 27 K and a bolometric luminosity of 5 \Lsun\ (Table~\ref{tbl-CrAbasic}).
The protostar is likely a binary, because is is a double radio source at 3.5 and
6 cm \citep{Choi08}.
. % Evans et al.  Tbol <  70 K Class 0;  70 <  Tbol <  650 Class I, 650  <  Tbol <  2800 K Class II
\subsubsection{IRS\,5 and IRS\,6}

IRS\,5 and IRS\,5\,N are Class I objects in  the western part  of the R CrA
dense cloud core. IRS\,5 is a binary with a separation of $\sim$ 0\ptsec6
\citep{Chen93,Peterson11}. It  was not detected at 1.3 mm with the SMA, although
IRS\,5\,N is quite a strong mm-source \citep{Peterson11}.  \citet{Nisini05}
estimated an extinction of 45 mag towards IRS\,5. This appears to be far too
high, as it would imply a luminosity of 200 \Lsun\ in the near-IR alone, which
is incompatible with the observed bolometric luminosity (Table~\ref{tbl-CrAbasic}). Our
estimate, 30 mag, is more in line with what one would expect, see
Figure~\ref{fig-Coronet_SEDs}. In the far infrared IRS\,5 and IRS\,5\,N show
spiral like structure surrounding the two stars (Figure~\ref{fig-FORCAST}). 
The bolometric temperatures derived by us (Table~\ref{tbl-CrAbasic}) are consistent
with their classification as Class I sources.

IRS\,6 is also a binary \citep{Nisini05} with a separation of 0\ptsec75.
\citet{Nisini05} found IRS\,6a to be an M2V star with an extinction of $\sim$ 30
mag. It is a weak X-ray source \citep{Forbrich06}. \citet{Peterson11} classify
it as a Class II object.  IRS\,6b is too faint to characterize. IRS\,6 is very
faint in the mid-IR.  We barely detected it with FORCAST (Table~\ref{tbl-CrAFORCAST}). In
the PACS image it looks like a diffuse extended object, i.e. all the far
infrared emission probably originates in a reflection nebula surrounding the
star.

\subsubsection{SMM\,2}

SMM\,2, also known as WBM\,55, is a Class I  source on the northeastern side of
the dense R CrA cloud core.  SMM\,2 is a strong sub-mm source
\citep{Groppi07,Peterson11}. It is relatively faint in  the FORCAST images and
somewhat stronger in the far infrared (Table~\ref{tbl-CrAFORCAST} \& \ref{tbl-CrAPACS}). 
Our results, Table~\ref{tbl-CrAbasic}, confirm that it is a Class I source. It has low
luminosity, 0.7 \Lsun, suggesting that it is  probably an early M-star.

%A two component graybody fit
%results in a dust temperature of 34 K ($\beta$ = 2.3 $\pm$ 0.4), resulting in a disk mass of
%%WDV
%0.18 \Msun, which is substantially larger than \citet{Peterson11} derived from their SMA
%observations at 226 GHz, 0.045 \Msun.

\subsection{The B\,59 cluster}

Our FORCAST observations, which targeted the central part of the B\,59 cluster,
detected seven sources, see Table~\ref{tbl-B59FORCAST}. From now on we abbreviate
the source name [BHB2007]~n as  BHB\,n, where n is the source number used by
\citet{Brooke07} and which we use in Tables~\ref{tbl-B59FORCAST},
\ref{tbl-B59PACS}, and \ref{tbl-B59SPIRE}, in order to  make the
paper easier to read. All of the sources, except BHB\,6, were previously
detected by MIPS \citep{Brooke07,Forbrich09}. Two of the sources: BHB\,10, and 
BHB\,11, were severely saturated in both the MIPS 24 and 70 $\mu$m images,
resulting in flux densities which were far too low. We just barely missed 
BHB\,15, which was at the very edge of our field, and which is only seen in a few
subframes at 19.7 $\mu$m. Of the sources detected by FORCAST, two are Class I
objects, one is a flat spectrum source, BHB\,7 (LkH$\alpha$\,346), and all the
rest are Class II \citep{Brooke07}. We did not detect BHB\,12, which was in our
field of view. It was rather faint at 24 $\mu$m (90 mJy), and it was not detected
with MIPS at 70 $\mu$m, nor was it detected by PACS.\footnote{We note that all 
the 20 candidate young stars reported by \citet{Brooke07} are real YSOs, except BHB\,5, which is
a background giant star \citep{Covey10}.}

\begin{figure}
\includegraphics[width=8.5cm, angle=0]{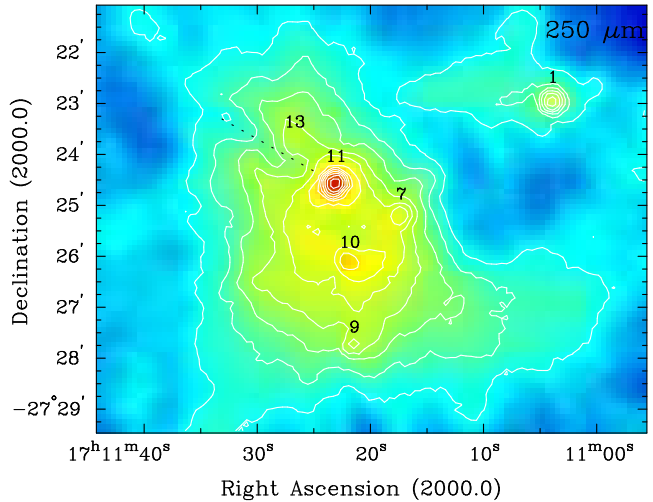}
\figcaption[]{
\label{fig-B59SPIRE}
A false color 250 $\mu$m SPIRE image of the B59 cloud plotted with a logarithmic
stretch and overlaid with white contours. The embedded sources seen in this
image are labeled with the source number from \citet{Brooke07}. The dotted line
outlines the cavity created by the outflow from BHB\,11 (B\,59\,MMS\,1).
 }   

\end{figure}

\begin{figure}[h]

\includegraphics[width=8.5cm, angle=0]{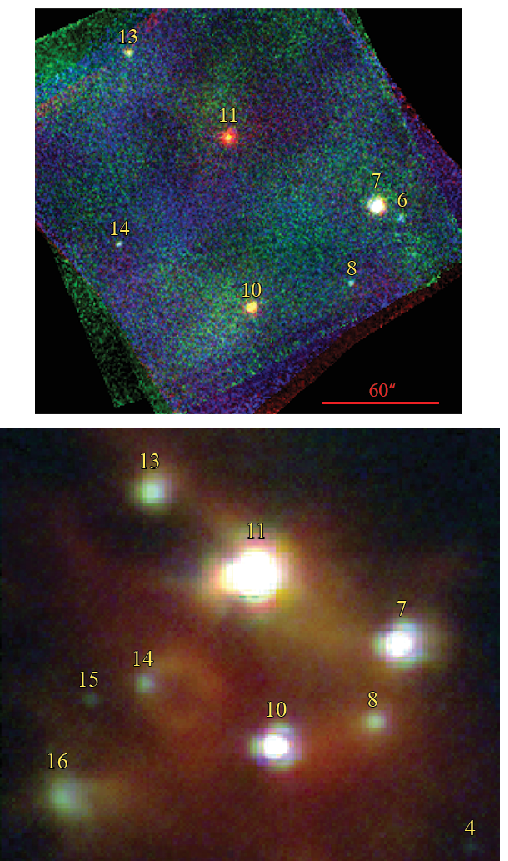}
\caption{{\it Top:}A three-color image of B\,59 with FORCAST; 11.1 $\mu$m (blue),
19.7 $\mu$m (green) and 31.5 $\mu$m (red). The very red star in the center of the
image is the deeply embedded Class I source BHB\,11. All the stars are
labeled. The red line at the bottom right shows the angular scale of the image,
60\arcsec. {\it Bottom:} PACS three-color image of approximately the same 
region of B\,59 as imaged with FORCAST.  Here the 70 $\mu$m filter is
coded red, 100  $\mu$m is blue, and the 160 $\mu$m filter is coded red. All the 
embedded B\,59 sources seen in the PACS color image are labeled.  BHB\,3 is
35\arcsec\ west of  BHB\,4 and outside the image.  The bubble seen SW of
BHB\,14 must have been created by this source .}
\label{fig-FORCASTPACS}
\end{figure}

Since PACS and SPIRE  cover the whole B\,59 cloud core, we have done photometry
of all sources believed to be part of the B\,59 cloud. Figure~\ref{fig-B59SPIRE}
shows a 250 $\mu$m SPIRE image of the whole B\,59 cloud. All the FORCAST
sources, except for BHB\,6, were detected by PACS, Table~\ref{tbl-B59PACS}. In
addition we pick up five sources, which were outside the area imaged by FORCAST
(Table~\ref{tbl-B59FORCAST}). Figure~\ref{fig-FORCASTPACS} shows a PACS color
image, which covers approximately the same area as FORCAST, but since the
FORCAST image was tilted relative to the sky reference frame, we pick up several
sources missed by FORCAST.  The only sources in the B\,59 core which are
missing are BHB\,3, BHB\,9 and BHB\,1. The latter is in a faint extended
filament NW of the cloud core.
Figure~\ref{fig-B59SPIRE} shows the whole cloud at 250 $\mu$m.  Six of the
sources are strong enough to be detected by SPIRE at 250 $\mu$m and 350 $\mu$m
(Table~\ref{tbl-B59SPIRE}). At 500 $\mu$m the emission is completely dominated
by the cloud and we were able to determine the flux density
of only the mm source MMS\,1, i.e., BHB\,11. Below we will discuss a selection of the
sources. For the rest we refer to Tables~\ref{tbl-B59FORCAST},
\ref{tbl-B59PACS}, and \ref{tbl-B59SPIRE} and Figures~\ref{fig-B59SPIRE} and
\ref{fig-FORCASTPACS}.

%2MASS J17111827-2725491 [BHB2007] 8   H mag 15.12, K 11.95
%\vskip 1cm
 
\begin{deluxetable*}{lllrrrl}[b]
%\tabletypesize{\scriptsize}
\tablecolumns{7}
%\tablenum{3}
\tablewidth{0pt} 
\tablecaption{Positions and flux densities of B\,59 sources observed with Herschel/PACS\label{tbl-B59PACS}}
\tablehead{
\colhead{Source} & \colhead{$\alpha$(2000.0)} & \colhead{$\delta$(2000.0)}  & \colhead{S(70 $\mu$m)}& \colhead{S(100 $\mu$m)} & \colhead{S(160 $\mu$m)} & \colhead{Comments}  \\ 
   & \colhead{[$^h$  $^m$ $^s$]}& \colhead{[$^\circ$ \arcmin\ \arcsec ]}& \colhead{[Jy]} & \colhead{[Jy]}  & \colhead{[Jy]} & \colhead{}
}
\startdata
$[$BHB2007$]$\,1 & 17 11 03.95 & $-$27 22 56.3  &  15.40 $\pm$  0.02 &  15.35 $\pm$ 0.01 & 12.32  $\pm$ 0.03 &\\
$[$BHB2007$]$\,3 & 17 11 11.74  & $-$27 26 55.0  & 0.28 $\pm$  0.02 &   0.26 $\pm$ 0.02 &  0.32 $\pm$ 0.02  &  \\
$[$BHB2007$]$\,4 & 17 11 14.47 & $-$27 26 57.0   &  0.12 $\pm$ 0.02 &    0.12 $\pm$ 0.01 &   $<$  0.04\phantom{aa1111} &   \\
$[$BHB2007$]$\,7 & 17 11 17.29 & $-$27 25 09.2   & 15.81 $\pm$ 0.03 & 15.54 $\pm$ 0.03 & 14.13 $\pm$ 0.03 &\\
$[$BHB2007$]$\,8 & 17 11 18.21 & $-$27 25 50.0   & 1.00 $\pm$ 0.02 & 1.30 $\pm$ 0.01 & 1.78 $\pm$ 0.07  & bright background\\
$[$BHB2007$]$\,9 & 17 11 21.51 & $-$27 27 42.6   & 2.88 $\pm$ 0.03 & 2.80 $\pm$ 0.01 & 3.78 $\pm$ 0.03 &\\
$[$BHB2007$]$\,10 & 17 11 22.15 & $-$27 26 03.0  & 14.58 $\pm$ 0.03 & 14.26 $\pm$ 0.02 & 13.00 $\pm$ 0.04 &\\
$[$BHB2007$]$\,11 & 17 11 23.09 & $-$27 24 33.2  & 82.75 $\pm$ 0.03 & 105.69 $\pm$ 0.03 & 108.60 $\pm$ 0.08 & B\,59-MMS\,1\\
$[$BHB2007$]$\,13 & 17 11 26.97  & $-$27 23 49.3 & 2.24 $\pm$ 0.03 & 3.15 $\pm$ 0.01 & 3.51 $\pm$ 0.04 &\\
$[$BHB2007$]$\,14 & 17 11 27.26  & $-$27 25 29.4 & 0.77 $\pm$ 0.03 & 1.07 $\pm$ 0.01 &  1.56 $\pm$ 0.03 & at apex of cavity\\
$[$BHB2007$]$\,15 & 17 11 29.42 & $-$27 25 37.9  & 0.16 $\pm$ 0.03 & 0.26 $\pm$ 0.01 & $<$ 0.07\phantom{aa1111} & \\
$[$BHB2007$]$\,16 & 17 11 30.40 & $-$27 26 29.4  & 1.17 $\pm$ 0.02 & 1.64  $\pm$ 0.02 & 1.13 $\pm$ 0.06 & nebulous
 \enddata
\end{deluxetable*}

\begin{deluxetable}{lrrc}[b]
%\tabletypesize{\scriptsize}
\tablecolumns{4}
\tablewidth{0pt} 
\tablecaption{Flux densities of B\,59 sources detected with Herschel/SPIRE\label{tbl-B59SPIRE}}
\tablehead{
\colhead{Source~~} & \colhead{S(250 $\mu$m)}& \colhead{S(350 $\mu$m)} & \colhead{S(500 $\mu$m)} \\ 
   & \colhead{[Jy]} & \colhead{[Jy]}  & \colhead{[Jy]} 
}
\startdata
$[$BHB2007$]$\,1  &  7.9 $\pm$  0.3 &  4.1  $\pm$ 0.1 & \nodata\\
$[$BHB2007$]$\,7 &  6.5  $\pm$ 1.2 &  2.1  $\pm$ 1.7 &  \nodata \\
$[$BHB2007$]$\,9  & 2.6 $\pm$ 0.6& 1.7 $\pm$ 0.6 &  \nodata  \\
$[$BHB2007$]$\,10 & 6.4  $\pm$ 2.3 & 3.4 $\pm$ 1.3 &   \nodata\\
$[$BHB2007$]$\,11 &  78.8 $\pm$ 3.2 & 42.9 $\pm$ 1.4 & 19.3 $\pm$ 1.7 \\
$[$BHB2007$]$\,13 &  1.9 $\pm$ 0.4  & 0.6 $\pm$ 0.3  &  \nodata  

 \enddata
\end{deluxetable}
\subsubsection{BHB\,1}
BHB\,1 is in the outskirt of the B\,59 cloud to the NW of the B\,59
cloud core (Figure~\ref{fig-B59SPIRE}). \citep{Brooke07} found it to be a flat
spectrum source, i.e.,  a star transitioning from Class I to Class II.  \citet{Covey10} assigned it a spectral type of  K 7. 
It was not covered by our FORCAST image. \citet{Reipurth93}
found it to be a binary  (LkH$\alpha$ 346\,NW) with a separation of  5\arcsec. This is the secondary.
The primary, LkH$\alpha$ 346\,SE, or BHB\,2, has a tertiary component with a separation 0\ptsec82  \citep{Chelli95,Zurlo21}. This
would make it a triple system if BHB\,1 and BHB\,2 AB are a bound system. Proper motion measurements by \citet{Zurlo21} found
BHB\,1 and BHB\,2 AB to have different proper motions, suggesting that it may not be a bound system.  \citet{Zurlo21} also found that the BHB\,2 binary
is more evolved than BHB\,1.
 BHB\,2 is much fainter in the mid-IR and it was not detected by MIPS at 24 $\mu$m. BHB\,1 has a circumstellar disk with a 
 large inner gap \citep{Alves20}. The disk, imaged with ALMA is seen almost
edge on. BHB\,2 was not detected by ALMA.

\subsubsection{BHB\,7}
BHB\,7  is the brightest source in  the  FORCAST images
(Table~\ref{tbl-B59FORCAST}, Figure~\ref{fig-FORCASTPACS}) and in the PACS bands
as well (Table~\ref{tbl-B59PACS}, Figure~\ref{fig-FORCASTPACS}), but rather
faint at 1.1 mm \citep{Hara13}. \citet{Brooke07} classify it as a
flat-spectrum source.  Based on
a medium resolution near-IR spectrum, \citet{Covey10} determined it to be a K5 star
with an A$_V$ of 16 mag. It is a cm radio source \citep{Dzib13}
and an X-ray source \citep{Forbrich10}. It is not known to drive an outflow. 

\subsubsection{BHB\,10}
BHB\,10 is a deeply embedded low luminosity Class I source \citep{Brooke07}. It
was not detected by 2MASS in JHK. It is not a cm-radio source, nor was it
detected as an X-ray source  \citep{Dzib13,Forbrich10}. It was saturated in MIPS
observations at 24 $\mu$m and 70 $\mu$m.  We adopt an extinction of 35 mag,
because it is in  a region of B\,59, where the extinction from background stars
is 46 mag \citep{Brooke07}. We determine  a bolometric temperature, T$_{bol}$,
of 110 K (Table~\ref{tbl-B59basic}), which is consistent with its classification as a
Class I source. \citet{Duarte12} argued that BHB\,10 drives a CO outflow.
Although it is plausible, the blue-shifted gas they observe could instead be
associated with BHB\,9 or BHB\,14.

\subsubsection{BHB\,11 (B\,59\,MMS\,1) }  
\label{sect-BHB11}
BHB\,11  was discovered as a sub-mm source by \citet{Reipurth96} who labeled it
B\,59\,MMS\,1. \citet{Brooke07} classified it as a Class 0/1 object, while
\citet{Riaz09}, who also detected it at J and H, classified it as a heavily
reddened Class I object with an A$_V$ = 19 - 30 mag. It is not detected with
FORCAST at 11 $\mu$m due to the very strong silicate absorption
(Table~\ref{tbl-B59FORCAST}), but it is the brightest source in the far infrared
(Figure~\ref{fig-B59SPIRE} and \ref{fig-FORCASTPACS}).  It is one of the
youngest members of the B\,59 cluster with an estimated age of 0.1 - 1 Myr
\citep{Riaz09}. It drives a bipolar molecular outflow
\citep{Riaz09,Duarte12,Hara13}. The NW outflow lobe has excavated the cloud,
which is seen as a narrow extended cavity in the SPIRE 250 and 350 $\mu$m images
(see Figure~\ref{fig-B59SPIRE}), suggesting the outflow is well collimated. The
PA for the NW outflow lobe is $\sim$ 57\degr. BHB\,11 is a binary with a
separation of 28 AU as seen by both Karl G. Jansky Very Large Array (VLA) and
Atacama Large Millimeter/Submillimeter Array (ALMA) \citep{Alves19}. BHB\,11A,
the northern source, is more massive  than its counterpart BHB\,11B (southern
member). \citet{Alves19} found that the two protostars are surrounded by dust
disks with radii of 3.1 $\pm$ 0.6 AU and 2.1 $\pm$ 0.5 AU with probably only a
few Jupiter masses and connected to a larger circumbinary disk through a complex
structure of  filaments. Most of the accretion  is toward the lower-mass
protostar.  BHB\,11  was saturated both at MIPS 24 $\mu$m and 70 $\mu$m. Using
FORCAST and PACS photometry, which are unsaturated, we are able to determine a
bolometric luminosity, L$_{bol}$  of 4.5 \Lsun\ (Table~\ref{tbl-B59basic}). The
bolometric temperature is 61.5 K for an assumed foreground extinction of 10 mag,
which puts BHB\,11 into the Class 0 regime. The foreground extinction would have
to be almost twice as high for the bolometric temperature to go over 70 K, the
boundary for a Class I source \citep{Chen95}. This sounds excessive, since most
of the extinction in this young source is likely to originate in the infalling
envelope.  Since the luminosity in the submillimeter, i.e. $\lambda \geq$  350
$\mu$m, is 0.16 $\pm$ 0.1 \Lsun, the ratio L$_{bol}$/L$_{submm}$ is in the range
17 - 75, which is well below 200, the upper limit for a Class 0 source. 
Although all the data suggest that BHB\,11 is a Class 0 source, it appears more evolved 
and is likely to be in transition to  a Class I object \citep{Riaz09,Alves19}. The boundary 
between Class 0 and Class I depends on luminosity \citep{Hatchell07}, and
for such a low luminosity system as BHB\,11, the bolometric temperature is probably
lower than it would be for a more luminous object at the same evolutionary stage.
A two component graybody fit  to the data in
this paper supplemented with  flux density at 1.3 mm from the SMA \citep{Hara13}
gives a dust emissivity index $\beta$ = 2.1 $\pm$ 0.1, a dust temperature of  22
-- 30 K, and a mass of 0.6 -- 0.8 \Msun, but the fit is not very well
constrained.  The warm dust component is $\sim$  50 K with a fitted size of
$\lesssim$ 2\arcsec. \citet{Riaz09} also found another low-luminosity ($\sim$
0.3 \Lsun{}) Class I object, 2M17112255, 8\arcsec\ from BHB\,11. This star is
potentially associated with  BHB\,11, which would make it a triple system.

\subsubsection{BHB\,13}
BHB\,13 was classified as a Class II object by \citep{Brooke07}, while
\citet{Forbrich09}, who used the spectral slope between K and 24 $\mu$m to
discriminate between Class I and Class II objects found it to be a 'flat
spectrum' source. We find it to be a Class II object (Table~\ref{tbl-B59basic}). It was
first identified as a binary (B\,59-1) by \citet{Reipurth93} with a separation
of 3\ptsec4. \citet{Koresko02} found it to be  a triple system, where the
tertiary  is separated by $\sim$ 0\ptsec1 from the primary. \citet{Covey10}
determined the primary to be a deeply embedded  M2 star with an A$_V$ of 12.5
mag from moderate resolution near IR spectra. This would make the star far too
bright in the visual. Here we adopt an extinction of 7 mag, which appears more
realistic. The star has both cm radio and X-ray emission
\citep{Dzib13,Forbrich10}.

\subsubsection{BHB\,14}
BHB\,14 was classified as a Class II object with $\sim$ 14 mag of extinction by
\citet{Brooke07}.  Our observations agree with this classification
(Table~\ref{tbl-B59basic}). \citet{Covey10} determined the spectral type K5  from
near-IR spectroscopy. It has not been detected as a cm radio source, nor does it
have X-ray emission. The PACS images (Figure~\ref{fig-FORCASTPACS}) show that it
sits at the apex of a large elliptical cavity with a length of $\sim$
100\arcsec\ (0.08 pc) at a PA of 248\degr, suggesting that it must have driven
an outflow in the past.
%\vskip 1 cm
\subsection{Non-detections}

We had expected to detect the Class 0 source SMM\,1\,C in the Coronet both with
FORCAST and PACS, but as discussed in Section~\ref{sect-SMM1C}, it is not
even seen at 70 $\mu$m, where the emission from IRS\,7\,A, 4\arcsec\ south of
it, still dominates. At 100 $\mu$m the emission peaks approximately halfway
between  SMM\,1\,C and IRS\,7\,A, while SMM\,1\,C is stronger at 160 $\mu$m
(Table~\ref{tbl-CrAPACS}). SMM\,1\,C is without doubt a deeply embedded Class 0
object. In the immediate vicinity of SMM\,1\,C \citet{Forbrich06} discovered
another two radio sources, FPM\,10 and FPM\,13, which have been detected with
the VLA both at 6.2 and 3.5 cm \citep{Choi08}. Both are extended and strongly
variable. FMP\,10 is probably associated with the outflow from  SMM\,1\,C
\citep{Choi08}, while the exciting source for FPM\,13 is unknown. 
\citet{Choi04} also found a 7 mm source CT\,3 halfway between IRS\,7\,A and
SMM\,1\,C, which is probably also outflow related \citep{Choi08}. Neither of
these sources has been detected in X-rays, nor were they seen with ALMA
\citep{Lindberg14}. The pre-stellar core  SMM\,1\,A , which dominates the dust
emission of the Coronet in the sub-millimeter \citep{Nutter05,Groppi07}, is not
detected by FORCAST, nor with PACS. \citet{Chen10} resolved the elongated core
into three dust condensations at 1.3 mm with the SMA. In the PACS wavebands the
emission peaks at the IRS\,7\,B and \,7\,A ridge (Figure~\ref{fig-FORCAST}).
There is still strong emission toward  SMM\,1\,A, but there is no enhancement
indicative of any embedded source. It is therefore still a pre-stellar core,
although it is very likely that some of the condensations in the core may
collapse and form stars.

\section{Cloud masses and star formation efficiency}

Since there are SPIRE images of both clouds we can use the thermal dust emission to
estimate the cloud masses over the same regions for which we have characterized the
YSO population. The SPIRE images processed  for extended emission are
zero-point corrected based on the {\it Planck}-HFI maps and recover extended
emission extremely well (see the SPIRE handbook\footnote{The SPIRE handbook is
available at \\ https://www.cosmos.esa.int/web/herschel/legacy-documentation}).
In that sense they are much better than SCUBA or Bolocam maps which always filter out
some of the extended emission.  We integrate over the cloud cores using the Starlink
Application GAIA, which allows us to integrate over the same area using the
same polygon for each filter. For the Coronet we integrated over  $\sim$
10 square arcminutes, corresponding to a cloud with a diameter  of  0.16 pc. For B\,59
the equivalent diameter of the cloud is 0.48 pc, i.e., the cloud core is about
three times as large as the Coronet one. We use the 250 $\mu$m and 350 $\mu$m images
to determine the dust temperature, T$_d$, see \citet{Indebetouw07}:

\[
T_d = \frac{hc/k(1/\lambda_2 - 1/\lambda_1)}{(3 + \beta)ln(\lambda_1/\lambda_2) +ln[F_\nu(\lambda_1)/F_\nu(\lambda_2)]}
\]

where h is the Planck constant, c is the speed of light, k is the Boltzmann
constant, $\lambda$ is the wavelength,  F$_\nu$ is the flux density and $\beta$
is the dust emissivity index. The frequency dependence of the dust emissivity,
$\kappa_\nu$, is normally expressed as a power law in the FIR, or  $\kappa_\nu
\propto  \nu^\beta$ (Hildebrand 1983,  who recommended  $\beta$ = 2 for
wavelengths $>$ 250 $\mu$m). This is consistent with observational constraints
on submillimeter dust emissivity \citep{Shirley11}, and frequently adopted in
analysis of {\it Herschel} data \citep[see e.g.][]{Roy14}. Here we also adopt
$\beta$ = 2, and note that the dust temperature  is not a strong function of
$\beta$. A dust emissivity index of 1.8 would change the dust temperature by
less than 10\% for the clouds we are studying. We first integrated fluxes for
the contribution from embedded sources to get the flux density for the cloud.
For the Coronet we do not have a good estimate, but extrapolating from the PACS
flux densities (Table~\ref{tbl-CrAPACS}) we end up with approximately 200 and 70 Jy
for 250 $\mu$m and 350 $\mu$m, respectively. Since the 250 $\mu$m emission from the cloud
is almost ten times higher, 1820 Jy,  the errors in our estimate of the
flux densities of the embedded sources are negligible. For B\,59 we can directly
sum up the flux densities in Table~\ref{tbl-B59SPIRE}. For the Coronet we derive
a dust temperature of 15.7 $\pm$ 0.3 K, while the temperature is 11.4 $\pm$ 0.2
K for B\,59. It is not surprising that the dust temperature is warmer in the
Coronet, since there are two HAEBE stars in the cloud and we clearly see warm
dust surrounding some of the embedded young stars (Figure~\ref{fig-FORCAST}).
Our derived dust temperature, 11.4 K, which is an average over the whole cloud,
agrees quite well with the average gas temperature, 10 - 12 K estimated
from CO(3--2) and isotopomers \citep{Duarte12} and  with a gas temperature of
11.3  $\pm$ 0.7 K derived from ammonia \citep{Rathborne09}. \citet{Redaelli17}
derived a somewhat higher dust temperature, 15 K, but they did not correct for
embedded sources and they spatially filtered their far infrared data.

Since the dust emission is largely optically thin, it is easy to derive the dust mass 
once we determine the dust temperature using  the formula

%\begin{equation}          % total mass calculation 
\[
M_{tot} = 1.9 \times 10^{-2} \biggl({1200\over\nu}\biggr)^{3+\beta}S_\nu
(e^{0.048\nu/T_d}-1)d^2
%\end{equation}
\]
\noindent
where d is the distance in $[$kpc$]$, and S$_\nu$ is the total flux at frequency
$\nu$ in $[$GHz$]$. M$_{\rm tot}$ is given in \Msun, see \citet{Sandell00}. In
this equation we have adopted the ``Hildebrand'' mass opacity, ${\rm \kappa_o}$,
defined at 250 $\mu$m (1200 GHz), i.e.  $\kappa_{\rm 1200GHz}$ = 0.1
cm$^{2}$g$^{-1}$ \citep{Hildebrand83} and a gas-to-dust ratio of 100. We
estimate the masses from the integrated flux densities at 250 $\mu$m. This
eliminates the frequency dependence of mass opacity. The only time we use the
dust emissivity index $\beta$, is when we determine the dust temperature, see
above. With these assumptions  we derive cloud masses of 28.0  $\pm$ 5.3 
\Msun\  and 68.5 $\pm$ 14.4 \Msun\ for the Coronet and
and B\,59 respectively, corresponding to average gas densities of
1.9 $\times 10^5$ cm$^{-3}$ and 1.7 $\times 10^4$ cm$^{-3}$. The dust mass opacity adopted by the {\it Herschel}
Gould Belt survey \citep{Roy14},  $\kappa_\lambda = 0.1 \times (\lambda/300
{\mu}m)^{-\beta}$ cm$^2$ g$^{-1}$, would increase the cloud masses by a factor
of 1.44, which is much higher than any  mass estimate obtained from
molecular line observations. \citet{Harju93} estimated 39 \Msun\ from
C$^{18}$O(1--0) of the area mapped by \citet{Taylor84}, which is more than two
times larger than the region we used for our mass estimate, i.e., within
errors it agrees with our estimate. For B\,59,
\citet{Hara13} derived a mass of 38 \Msun\ from their  AzTEC 1.1mm map, but
millimeter observations filter out all of the low density cirrus emission, so
one would expect the mass to be on the low side.   \citet{Duarte12} using
C$^{18}$O(3--2) estimate 47 \Msun\ integrated over a larger area than what we
use. However, this estimate may be on the low side, since they consistently find
much higher masses from  analyzing  $^{13}$CO(3--2) than what they get from
C$^{18}$O(3--2). Therefore, considering the uncertainties in these estimates, our
results compare reasonably well with previous values.

 Since we estimated the masses of the cloud cores and since we have identified
 the embedded young stars in each core, we can now estimate the star formation
 efficiency. In the Coronet the two HAEBE stars R CrA and T Cr A dominate the
 mass, most of the other stars are low mass mid-K to M stars  with masses
 of $\sim$ 0.5 \Msun to $<$ 0.1 \Msun\ \citep{Nisini05}.  We estimate the total
 mass of the Coronet YSOs to be $\sim$ 9.1 $\pm$ 1.2  \Msun. For B\,59 we
 use the mass estimates by \citet{Covey10}, i.e., 9 - 13 \Msun, depending on which evolutionary tracks
 they used.  We
 therefore get star formation efficiencies of 27 $\pm$ 7 \% and 14 $\pm$ 5 \% for  the Coronet and
 B\,59, respectively. It is well known that star formation is a function of
 cloud density. \citet{Kennicutt98} showed that the star formation rate in a
 large sample of spiral and starburst galaxies follows a power law with an
 exponent of 1.4, the ``Kennicutt--Schmidt law''. The same is seen in our
 Galaxy.  However, this relation appears to be even steeper for local molecular
 clouds  \citep{Heiderman10}. It therefore makes sense that the star
 formation efficiency is much higher in the Coronet than in B\,59, since we have
 shown that the gas density is about ten times higher in the Coronet than in
 B\,59. Nor is it surprising that the star formation efficiencies in these
 dense cores are much  higher than the global star formation rates of the large
 cloud complexes where they reside. For  the whole CrA cloud complex,
 \citet{Peterson11} estimate $\sim$ 8~\%, while \citet{Forbrich09} finds  $\sim$
 6~\% for the entire Pipe nebula. Even so, these star formation rates are
 relatively high compared to most star forming regions, which have star
 formation efficiencies in the range 3 -- 6~\% \citep{Evans09}.

\section{Summary and Conclusion}

We present mid infrared imaging of two young clusters, the Coronet in the CrA
cloud core and B\,59 in the Pipe Nebula,  using the FORCAST camera on the
Stratospheric Observatory for Infrared Astronomy. The Coronet, at a distance of
154 pc, is part of the large CrA cloud complex, while B\,59 in the Pipe nebula
is at a distance of 163 pc. We also analyze {\it  Herschel} Space Observatory
PACS and SPIRE images of the same clouds obtained from the  {\it  Herschel}
Science archive. Most of the  {\it  Herschel} data have never been published
before. \citet{Sicilia13} analyzed some of the {\it Herschel} PACS/SPIRE images
of the Coronet, but they  only discussed photometry obtained by
fast scanning at 100 and 160 $\mu$m. Here we analyze photometry for all three
bands, most of which has been obtained in small map mode, which has better
angular resolution.  Photometry of B\,59 with PACS and SPIRE has not been
previously published.

Both clusters are nearby and at very similar distances.  The Coronet cluster is
younger, 0.5 -- 1 Myr \citep{Sicilia08} than B\,59, which has an age of $\sim$
2.6 Myr \citep{Covey10}. Most of the embedded sources in the Coronet are Class I
objects while most of the stars in B59 are Class II or Flat spectrum sources,
which is consistent with B\,59 being older. Both clusters, however, have at least
one Class 0 object, which indicates that star formation is still ongoing in both
clusters.

Our FORCAST imaging did not reveal any new sources neither in the Coronet not
B\,59. This is largely because both clusters are nearby and
have been subject to many observational studies. However, our observations
provide much more accurate values for fundamental properties like bolometric
luminosity, which has been poorly known, especially in the Coronet, see
Table~\ref{tbl-CrAbasic}. Our observations generally agree with the evolutionary
classes of the embedded sources determined in previous studies. Our results for
B\,59 agree better with \citet{Brooke07} than with the evolutionary classes
assigned by \citet{Forbrich09}, because the spectral indices used by
\citeauthor{Forbrich09} were computed without extinction correction. We find
BHB\,11 in B59 to be a Class 0 source, because the bolometric temperature
that we determine, 61.5  K, is well within the range for Class 0 sources, $<$ 70
K \citep{Myers93,Chen95}, probably in transition to Class I.

Using SPIRE images we determine the physical properties of the Coronet and B\,59
cloud cores.  We find that the Coronet has a size of $\sim$ 0.16 pc, a dust
temperature of 15.7 K and a mass of 28 \Msun, corresponding to an average gas
density of 1.9 $\times 10^5$ cm$^{-3}$. The B\,59 is about twice as large but
colder and much less dense, with a size of  0.48 pc, a dust
temperature of 11.4 K, and a mass of 68.5 \Msun, corresponding to a gas density
of 1.7 $\times 10^4$ cm$^{-3}$.

We determine that the  star formation efficiency is about  twice as high in the dense
 and compact Coronet cloud core ($\sim$ 27~\%) compared to B\,59 ($\sim$ 14~\%),
which is more extended and less dense. In both cloud complexes these clusters
represent the most active star forming part of the cloud. The star formation
efficiency of the whole Cr A cloud complex, which spans over about 1.5\degr, is
only about  $\sim$ 8~\% \citep{Peterson11}, while it is  $\sim$ 6~\% for the
entire Pipe nebula \citep{Forbrich09}.

\acknowledgements 
A thorough reading of 
the paper by an anonymous referee helped us to improve the presentation considerably.
Based in part on observations made with the NASA/DLR Stratospheric Observatory
for Infrared Astronomy (SOFIA). SOFIA is jointly operated by the Universities
Space Research Association, Inc. (USRA), under NASA contract NNA17BF53C, and the
Deutsches SOFIA Institut (DSI) under DLR contract 50 OK 0901 to the University
of Stuttgart. This work is also in part based on {\it Herschel} PACS and SPIRE
data. PACS has been
developed by a consortium of institutes led by MPE (Germany)
and includingUVIE (Austria); KU Leuven, CSL, IMEC (Belgium); CEA, LAM
(France); MPIA (Germany); INAF-IFSI/OAA/OAP/OAT, LENS, SISSA (Italy);
IAC (Spain). This development has been supported by the funding agencies
BMVIT (Austria), ESA-PRODEX (Belgium), CEA/CNES (France), DLR (Germany),
ASI/INAF (Italy), and CICYT/MCYT (Spain). SPIRE has been developed by
a consortium of institutes led by Cardiff University (UK) and including Univ.
Lethbridge (Canada); NAOC (China); CEA, LAM (France); IFSI, Univ. Padua (Italy);
IAC (Spain); Stockholm Observatory (Sweden); Imperial College London, RAL,
UCL-MSSL, UKATC, Univ. Sussex (UK); and Caltech, JPL, NHSC, Univ. Colorado
(USA). This development has been supported by national funding agencies: CSA
(Canada); NAOC (China); CEA, CNES, CNRS (France); ASI (Italy); MCINN (Spain);
SNSB (Sweden); STFC, UKSA (UK); and NASA (USA). Financial support for this
research was provided by NASA through grant  SOF 07-0045, issued by USRA.
We thank Dr. A. Helton for help in planning the observations.

\newpage

\appendix
\renewcommand\thefigure{\thesection.\arabic{figure}}   
\section{Appendix A: SED plots}
\setcounter{figure}{0} 
Below we present SEDs from J band (1.42 $\mu$m) to 1.3 mm of all the observed sources  in the Coronet (Figure A.1) and B\,59 (Figure A.2).

\begin{figure*}[ht]
\includegraphics[width=\textwidth, angle=0]{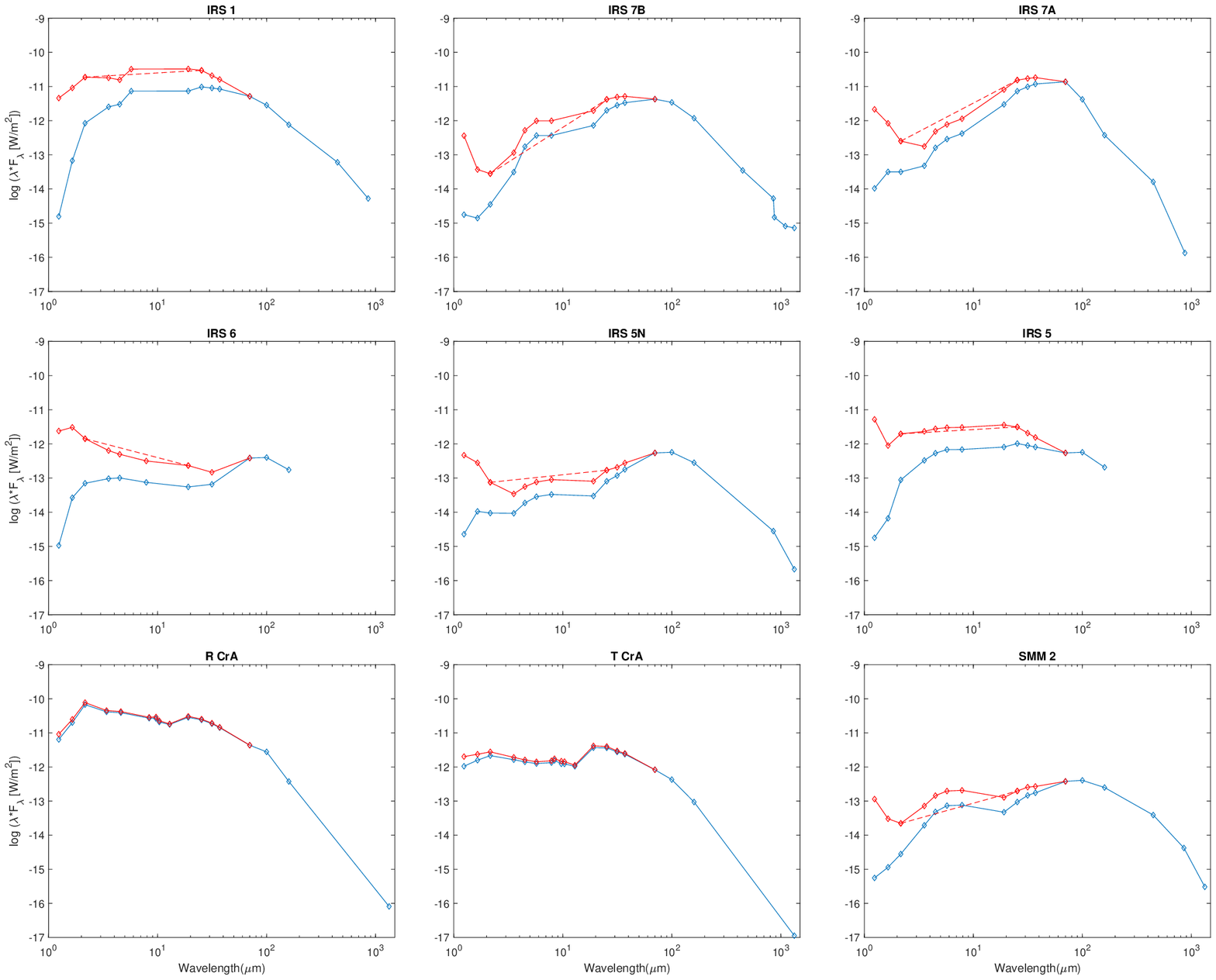}
\figcaption[]{SEDs of all the young stars in the Coronet. We have supplemented our FORCAST and PACS photometry with 2MASS data and IRAC data \citep{Peterson11},
together with published ground based mid infrared photometry   \citep{Wilking86,Berilli92,Prusti94}  and millimeter and submillimeter data  \citep{Groppi07,Peterson11,Lindberg14}.
The blue curve shows the observed data points, the red curve has been corrected for the extinction given in Table~\ref{tbl-CrAbasic}. The dashed red curve shows the slope of the 
spectral index, $\alpha^*$, from K$_s$ to 19.7 $\mu$m.
\label{fig-Coronet_SEDs}
}
\end{figure*}

\begin{figure*}[ht]
\includegraphics[width=\textwidth, angle=0]{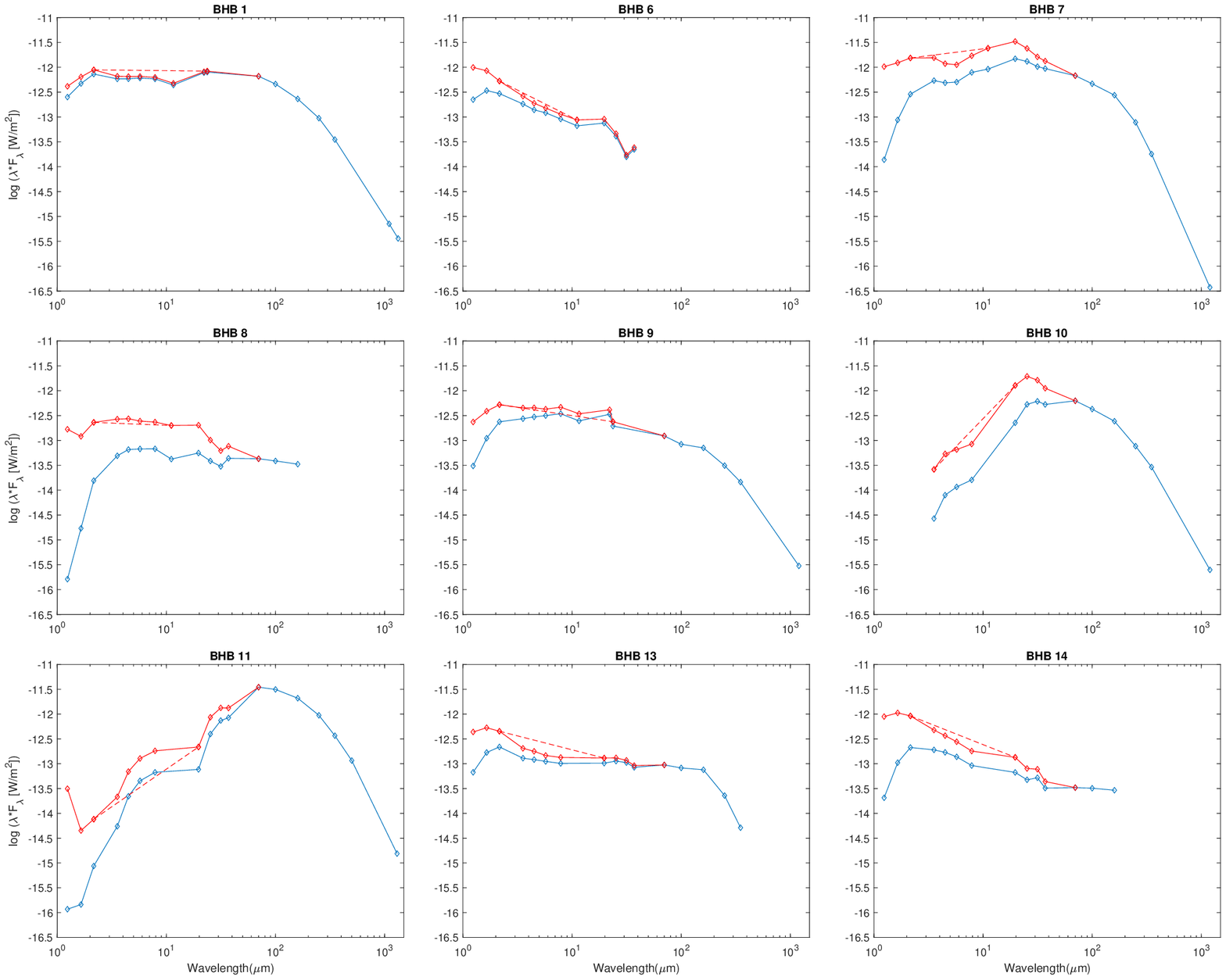}
\figcaption[]{SEDs of all the young stars in B\,59. We have supplemented our FORCAST, PACS and SPIRE photometry with 2MASS, IRAC and MIPS data \citep{Brooke07,Forbrich09} and
for a few sources also with WISE photometry in band W3 and W4. Millimeter and submillimeter photometry comes from  \citet{Roman12,Hara13,Alves20}. J and H photometry for BHB\,11 is from \citet{Riaz09}.
The SEDs are plotted the same way as in Figure A.1. The extinction used to correct the data is given in Table~\ref{tbl-B59basic}. The spectral index, $\alpha^*$, is for most sources plotted from K$_s$ to 11.1 $\mu$m.
For BHB\,10, which was not detected by 2MASS, the starting point is 3.6 $\mu$m. For BHB\,1 and BHB\,11, which were not covered by the FORCAST image, the endpoint is MIPS 24 $\mu$m.
\label{fig-B59_SEDs}
}
\end{figure*}

\end{document}